# Electromagnetic Stress Tensor in Ponderable Media

**Masud Mansuripur**

College of Optical Sciences, The University of Arizona, Tucson, Arizona 85721
masud@optics.arizona.edu

**Abstract**: We derive an expression for the Maxwell stress tensor in a magnetic dielectric medium specified by its permittivity $\varepsilon$ and permeability $\mu$. The derivation proceeds from the generalized form of the Lorentz law, which specifies the force exerted by the electromagnetic $\boldsymbol{E}$ and $\boldsymbol{H}$ fields on the polarization $\boldsymbol{P}$ and magnetization $\boldsymbol{M}$ of ponderable media. Our stress tensor differs from the well-known tensors of Abraham and Minkowski, which have been at the center of a century-old controversy surrounding the momentum of the electromagnetic field in transparent materials.

**OCIS codes**: (260.2110) Electromagnetic theory; (140.7010) Trapping.

## References and links

## 1. Introduction

In the classical theory of electromagnetism, the Poynting vector $\boldsymbol{S}(\boldsymbol{r}, t) = \boldsymbol{E} \times \boldsymbol{H}$ specifies the rate of flow of energy at a given location in space $\boldsymbol{r}$ and time $t$. The integral of $\boldsymbol{S}(\boldsymbol{r}, t)$ over a closed surface is equal to the rate of increase of stored energy, $\partial \mathcal{E}/\partial t$, within the volume enclosed by the surface, plus the rate of loss (or minus the rate of gain) of energy throughout that volume [1]. Similarly, the stress tensor $\boldsymbol{T}(\boldsymbol{r}, t)$ specifies the rate of flow of momentum (i.e., momentum crossing unit area per unit time) at a given point in space and time.

The electromagnetic stress tensor in the free space, known as Maxwell's stress tensor, is readily derived from the Lorentz law of force in conjunction with the microscopic form of Maxwell's equations [1]. However, in ponderable media where polarization density $\boldsymbol{P}$ and magnetization density $\boldsymbol{M}$ describe the electromagnetic properties of the material (and where the macroscopic version of Maxwell's equations, incorporating $\boldsymbol{P}$ and $\boldsymbol{M}$, are applicable), the form of the stress tensor has been the subject of debate and controversy for the past century [2]. In particular, the tensors of Minkowski and Abraham have been subjected to endless analysis; experiments have been designed to confirm or refute their corresponding predictions, and each tensor, while successful in some respects, has been found inadequate in others.

In a series of recent publications, we have shown that the pressure and momentum of the electromagnetic field can be obtained directly from the Lorentz law of force in conjunction with macroscopic Maxwell equations, without the need to determine the stress tensor. However, like Poynting's vector, stress tensor is a powerful tool that can be used to calculate forces and momenta by employing the knowledge of the fields at the boundaries, without having to pay attention to the details of what goes on inside a volume. In this paper we derive a general expression for the electromagnetic stress tensor inside ponderable media and show that, in cases that we have studied previously, the new tensor reproduces the old results.

## 2. Electromagnetic stress tensor in ponderable media

In a recent publication [3] we derived the following generalized expression for the Lorentz force density in a linear isotropic medium specified by its $\mu$ and $\varepsilon$ parameters:

$$\boldsymbol{F}(\boldsymbol{r},t) = (\boldsymbol{P}\cdot\nabla)\boldsymbol{E} + (\boldsymbol{M}\cdot\nabla)\boldsymbol{H} + (\partial\boldsymbol{P}/\partial t)\times\mu_o\boldsymbol{H} - (\partial\boldsymbol{M}/\partial t)\times\varepsilon_o\boldsymbol{E}. \tag{1}$$

In conjunction with Eq. (1), Maxwell's equations in the MKSA system of units are:

$$\nabla\cdot\boldsymbol{D} = \rho_{\text{free}}, \tag{2a}$$

$$\nabla\times\boldsymbol{H} = \boldsymbol{J}_{\text{free}} + \partial\boldsymbol{D}/\partial t, \tag{2b}$$

$$\nabla\times\boldsymbol{E} = -\partial\boldsymbol{B}/\partial t, \tag{2c}$$

$$\nabla\cdot\boldsymbol{B} = 0. \tag{2d}$$

In what follows, the medium will be assumed to have neither free charges nor free currents, that is, $\rho_{\text{free}} = 0$ and $\boldsymbol{J}_{\text{free}} = 0$. In the above equations, the electric displacement $\boldsymbol{D}$ and the magnetic induction $\boldsymbol{B}$ are related to the polarization density $\boldsymbol{P}$ and the magnetization density $\boldsymbol{M}$ as follows:

$$\boldsymbol{D} = \varepsilon_o\boldsymbol{E} + \boldsymbol{P} = \varepsilon_o(1+\chi_e)\boldsymbol{E} = \varepsilon_o\,\varepsilon\boldsymbol{E}, \tag{3a}$$

$$\boldsymbol{B} = \mu_o\boldsymbol{H} + \boldsymbol{M} = \mu_o(1+\chi_m)\boldsymbol{H} = \mu_o\,\mu\boldsymbol{H}. \tag{3b}$$

Using different arguments, Hansen and Yaghjian [4] have arrived at the same expression as Eq. (1) for the Lorentz force under quite general conditions. Also, Kemp *et al*, in their analysis of momentum in left-handed media [5], use essentially the same force equation. Our derivation of Eq. (1) in [3] started from the well-known Lorentz formula, $\boldsymbol{F} = q(\boldsymbol{E}+\boldsymbol{V}\times\boldsymbol{B})$, but it soon became apparent that magnetic dipoles cannot be treated as simple Amperian current loops; conservation of momentum demanded certain modifications of the original Lorentz law. In particular, a new term had to be introduced to account for the force experienced by magnetic dipoles. Rather than attempting to justify Eq. (1) on the basis of the original Lorentz law, we believe that one should simply accept it as a law of nature, on par with Maxwell's equations. Not only are these five equations consistent among themselves, they also comply with the laws of energy and momentum conservation.

The first term on the right hand side of Eq. (1) may be rewritten using the identities:

$$(\boldsymbol{P}\cdot\nabla)\boldsymbol{E} + (\nabla\cdot\boldsymbol{P})\boldsymbol{E} = \partial(P_x\boldsymbol{E})/\partial x + \partial(P_y\boldsymbol{E})/\partial y + \partial(P_z\boldsymbol{E})/\partial z, \tag{4a}$$

$$\nabla\cdot\boldsymbol{P} = \nabla\cdot(\boldsymbol{D} - \varepsilon_o\boldsymbol{E}) = \rho_{\text{free}} - \varepsilon_o\nabla\cdot\boldsymbol{E} = -\,\varepsilon_o\nabla\cdot\boldsymbol{E}. \tag{4b}$$

A similar treatment can be applied to the second term in Eq. (1); here $\nabla\cdot\boldsymbol{B}$ is readily set to zero in accordance with Maxwell's $4^{\text{th}}$ equation.

$$(\boldsymbol{M}\cdot\nabla)\boldsymbol{H} + (\nabla\cdot\boldsymbol{M})\boldsymbol{H} = \partial(M_x\boldsymbol{H})/\partial x + \partial(M_y\boldsymbol{H})/\partial y + \partial(M_z\boldsymbol{H})/\partial z. \tag{5a}$$

$$\nabla\cdot\boldsymbol{M} = \nabla\cdot(\boldsymbol{B} - \mu_o\boldsymbol{H}) = \nabla\cdot\boldsymbol{B} - \mu_o\nabla\cdot\boldsymbol{H} = -\,\mu_o\nabla\cdot\boldsymbol{H}. \tag{5b}$$

The third term in Eq. (1) is rewritten by substituting for $\boldsymbol{P}$ in terms of $\boldsymbol{D}$ and $\boldsymbol{E}$, then invoking Maxwell's $2^{\text{nd}}$ equation. Similarly, the fourth term is rewritten by substituting for $\boldsymbol{M}$ in terms of $\boldsymbol{B}$ and $\boldsymbol{H}$, then invoking Maxwell's $3^{\text{rd}}$ equation. We find

$$(\partial\boldsymbol{P}/\partial t)\times\mu_o\boldsymbol{H} = (\partial\boldsymbol{D}/\partial t - \varepsilon_o\partial\boldsymbol{E}/\partial t)\times\mu_o\boldsymbol{H} = \mu_o(\nabla\times\boldsymbol{H})\times\boldsymbol{H} - \varepsilon_o\mu_o(\partial\boldsymbol{E}/\partial t)\times\boldsymbol{H}, \tag{6a}$$

$$(\partial\boldsymbol{M}/\partial t)\times\varepsilon_o\boldsymbol{E} = (\partial\boldsymbol{B}/\partial t - \mu_o\partial\boldsymbol{H}/\partial t)\times\varepsilon_o\boldsymbol{E} = -\varepsilon_o(\nabla\times\boldsymbol{E})\times\boldsymbol{E} - \varepsilon_o\mu_o(\partial\boldsymbol{H}/\partial t)\times\boldsymbol{E}. \tag{6b}$$

Substitution from Eqs. (4-6) into Eq. (1), followed by rearranging and combining the various terms yields,

$$\boldsymbol{F}(\boldsymbol{r},t) = (\partial/\partial x)(P_x\boldsymbol{E} + M_x\boldsymbol{H}) + (\partial/\partial y)(P_y\boldsymbol{E} + M_y\boldsymbol{H}) + (\partial/\partial z)(P_z\boldsymbol{E} + M_z\boldsymbol{H})$$
$$+ \varepsilon_o[(\nabla\cdot\boldsymbol{E})\boldsymbol{E} + (\nabla\times\boldsymbol{E})\times\boldsymbol{E}] + \mu_o[(\nabla\cdot\boldsymbol{H})\boldsymbol{H} + (\nabla\times\boldsymbol{H})\times\boldsymbol{H}] - \varepsilon_o\mu_o\partial(\boldsymbol{E}\times\boldsymbol{H})/\partial t. \quad (7)$$

This equation can be further expanded and rearranged to yield,

$$\boldsymbol{F}(\boldsymbol{r},t) + (\partial/\partial t)(\boldsymbol{E}\times\boldsymbol{H}/c^2) = (\partial/\partial x)\{P_x\boldsymbol{E} + M_x\boldsymbol{H} + \varepsilon_o[\tfrac{1}{2}(E_x^2-E_y^2-E_z^2)\hat{\boldsymbol{x}}+E_xE_y\hat{\boldsymbol{y}}+E_xE_z\hat{\boldsymbol{z}}]$$
$$+ \mu_o[\tfrac{1}{2}(H_x^2-H_y^2-H_z^2)\hat{\boldsymbol{x}}+H_xH_y\hat{\boldsymbol{y}}+H_xH_z\hat{\boldsymbol{z}}]\}$$
$$+ (\partial/\partial y)\{P_y\boldsymbol{E} + M_y\boldsymbol{H} + \varepsilon_o[E_xE_y\hat{\boldsymbol{x}}+\tfrac{1}{2}(E_y^2-E_x^2-E_z^2)\hat{\boldsymbol{y}}+E_yE_z\hat{\boldsymbol{z}}]$$
$$+ \mu_o[H_xH_y\hat{\boldsymbol{x}}+\tfrac{1}{2}(H_y^2-H_x^2-H_z^2)\hat{\boldsymbol{y}}+H_yH_z\hat{\boldsymbol{z}}]\}$$
$$+ (\partial/\partial z)\{P_z\boldsymbol{E} + M_z\boldsymbol{H} + \varepsilon_o[E_xE_z\hat{\boldsymbol{x}}+E_yE_z\hat{\boldsymbol{y}}+\tfrac{1}{2}(E_z^2-E_x^2-E_y^2)\hat{\boldsymbol{z}}]$$
$$+ \mu_o[H_xH_z\hat{\boldsymbol{x}}+H_yH_z\hat{\boldsymbol{y}}+\tfrac{1}{2}(H_z^2-H_x^2-H_y^2)\hat{\boldsymbol{z}}]\}. \quad (8)$$

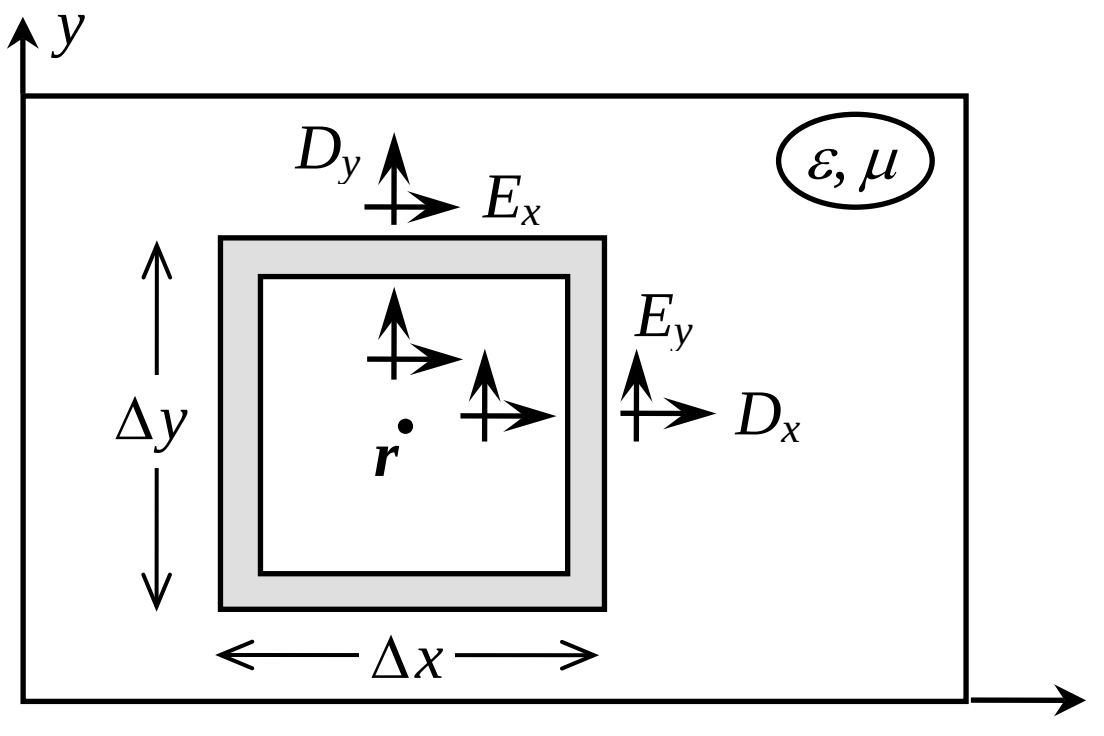

Fig. 1. A small cube of dimensions $\Delta x\times\Delta y\times\Delta z$ within a magnetic dielectric is separated from the surrounding medium by a fictitious vacuum-filled gap; the medium is specified by its ($\varepsilon$, $\mu$) parameters. Assuming the gap is sufficiently narrow (compared to the wavelength of the electromagnetic field), its presence should not affect the distribution of the fields throughout the medium. Within the gap, however, the various components of the electromagnetic field are determined by the standard boundary conditions derived from Maxwell's equations. In general, the tangential components of ***E*** and ***H*** remain continuous across the gap, while, in the perpendicular direction, the components of ***D*** and ***B*** retain continuity.

Next, we integrate Eq. (8) over the small $\Delta x\times\Delta y\times\Delta z$ cube depicted in Fig. 1, normalize the resultant by the cube's volume, and consider the limit when ($\Delta x$, $\Delta y$, $\Delta z$)$\rightarrow$0. The left-hand side of Eq. (8) thus remains intact, but several changes occur on the right-hand side. For instance, in the first term, integration over $x$ yields the argument of $\partial/\partial x$, evaluated in the gaps on the left- and right-hand sides of the cube, then subtracted from each other. In these gaps, $P_x = 0$, $M_x = 0$, $\varepsilon_o E_x = D_x$, and $\mu_o H_x = B_x$, while the remaining components of ***E*** and ***H*** retain the values that they have in the adjacent material environment. (These gap fields are found by invoking standard boundary conditions, namely, the continuity of tangential ***E*** and ***H***, as well as perpendicular ***D*** and ***B*** components.) Similar arguments apply to the second and third terms on the right-hand side of Eq. (8), provided that, in the case of the 2nd (3rd) term, the initial integration is carried out over $y$ ($z$). When the integrals are fully evaluated and the result is normalized by the volume of the cube, we find, in the limit of a vanishing cube,

$$\boldsymbol{F}(\boldsymbol{r},t) + \frac{\partial}{\partial t}(\boldsymbol{E}\times\boldsymbol{H}/c^2) = \frac{\partial}{\partial x}\{[\tfrac{1}{2}(\varepsilon_o^{-1}D_x^2-\varepsilon_oE_y^2-\varepsilon_oE_z^2)\hat{\boldsymbol{x}}+D_xE_y\hat{\boldsymbol{y}}+D_xE_z\hat{\boldsymbol{z}}] \quad (9)$$
$$+ [\tfrac{1}{2}(\mu_o^{-1}B_x^2-\mu_oH_y^2-\mu_oH_z^2)\hat{\boldsymbol{x}}+B_xH_y\hat{\boldsymbol{y}}+B_xH_z\hat{\boldsymbol{z}}]\} \quad \text{continued} \cdots$$

$$+\frac{\partial}{\partial y}\{[E_xD_y\hat{\boldsymbol{x}}+½(\varepsilon_o^{-1}D_y^2-\varepsilon_oE_x^2-\varepsilon_oE_z^2)\hat{\boldsymbol{y}}+D_yE_z\hat{\boldsymbol{z}}]+[H_xB_y\hat{\boldsymbol{x}}+½(\mu_o^{-1}B_y^2-\mu_oH_x^2-\mu_oH_z^2)\hat{\boldsymbol{y}}+B_yH_z\hat{\boldsymbol{z}}]\}$$

$$+\frac{\partial}{\partial z}\{[E_xD_z\hat{\boldsymbol{x}}+E_yD_z\hat{\boldsymbol{y}}+½(\varepsilon_o^{-1}D_z^2-\varepsilon_oE_x^2-\varepsilon_oE_y^2)\hat{\boldsymbol{z}}]+[H_xB_z\hat{\boldsymbol{x}}+H_yB_z\hat{\boldsymbol{y}}+½(\mu_o^{-1}B_z^2-\mu_oH_x^2-\mu_oH_y^2)\hat{\boldsymbol{z}}]\}.$$

Equation (9) clearly identifies the Abraham momentum density $\boldsymbol{E}\times\boldsymbol{H}/c^2$ as the electromagnetic momentum density $\boldsymbol{G}(\boldsymbol{r}, t)$, and yields the following stress tensor $T_{ij}$ (i.e., rate of flow of momentum per unit area per unit time) within the medium:

$$T_{xx} = ½(\varepsilon_oE_y^2+\varepsilon_oE_z^2-\varepsilon_o^{-1}D_x^2) + ½(\mu_oH_y^2+\mu_oH_z^2-\mu_o^{-1}B_x^2), \tag{10a}$$

$$T_{yx} = -D_xE_y-B_xH_y, \tag{10b}$$

$$T_{zx} = -D_xE_z-B_xH_z, \tag{10c}$$

$$T_{xy} = -E_xD_y-H_xB_y, \tag{10d}$$

$$T_{yy} = ½(\varepsilon_oE_x^2+\varepsilon_oE_z^2-\varepsilon_o^{-1}D_y^2) + ½(\mu_oH_x^2+\mu_oH_z^2-\mu_o^{-1}B_y^2), \tag{10e}$$

$$T_{zy} = -D_yE_z-B_yH_z, \tag{10f}$$

$$T_{xz} = -E_xD_z-H_xB_z, \tag{10g}$$

$$T_{yz} = -E_yD_z-H_yB_z, \tag{10h}$$

$$T_{zz} = ½(\varepsilon_oE_x^2+\varepsilon_oE_y^2-\varepsilon_o^{-1}D_z^2) + ½(\mu_oH_x^2+\mu_oH_y^2-\mu_o^{-1}B_z^2). \tag{10i}$$

Equation (9) may thus be written as the following streamlined expression of momentum conservation:

$$\nabla\cdot\boldsymbol{T}+\boldsymbol{F}(\boldsymbol{r},t)+\partial\boldsymbol{G}(\boldsymbol{r},t)/\partial t = 0. \tag{11}$$

In its specific combination of the various components of the $\boldsymbol{E}$, $\boldsymbol{D}$, $\boldsymbol{H}$ and $\boldsymbol{B}$ fields, the stress tensor of Eq. (10) differs from both Abraham and Minkowski tensors. A similar (although by no means identical) tensor has been derived by Yaghjian [6], who has advocated methods of analysis that in many respects resemble our methods.

**Example 1**. A plane electromagnetic wave propagates along the $z$-axis inside a medium specified by its $(\varepsilon, \mu)$ parameters. The linearly polarized plane-wave has $E$-field amplitude $E_o\hat{\boldsymbol{x}}$ and $H$-field amplitude $H_o\hat{\boldsymbol{y}} = Z_o^{-1}\sqrt{\varepsilon/\mu}\,E_o\hat{\boldsymbol{y}}$, where $Z_o=\sqrt{\mu_o/\varepsilon_o}$ is the impedance of the free space. Assuming a monochromatic plane wave with angular frequency $\omega$, the rate of flow of momentum (per unit area per unit time) along the $z$-axis will be given by Eq. (10i) as follows:

$$T_{zz} = ½(\varepsilon_oE_x^2+\mu_oH_y^2) = ½\,\varepsilon_o[1+(\varepsilon/\mu)]E_o^2\cos^2(\omega t). \tag{12}$$

The rate of flow of energy (per unit area per unit time) is the Poynting vector, $\boldsymbol{S}=\boldsymbol{E}\times\boldsymbol{H}$. Therefore,

$$S_z = E_xH_y = Z_o^{-1}\sqrt{\varepsilon/\mu}\,E_o^2\cos^2(\omega t). \tag{13}$$

Suppose a total of $N$ photons cross the $xy$-plane at $z=0$ during the time interval $[0,\tau]$. Since each photon has energy $hf$, we have $<S_z>\tau = Nhf$ (angled brackets denote time-averaging). The total momentum crossing the same plane during the same time interval will therefore be $<T_{zz}>\tau = ½N(\sqrt{\varepsilon/\mu}+\sqrt{\mu/\varepsilon})hf/c$. A single photon's momentum is thus $½(\sqrt{\varepsilon/\mu}+\sqrt{\mu/\varepsilon})hf/c$, in agreement with our previous results [3]. This photon momentum, which in non-magnetic dielectrics is equal to the arithmetic average of the Minkowski and Abraham momenta, is always greater than the photon momentum in free space, $hf/c$. In general, the photon momentum consists of an electromagnetic part and a mechanical part. In a non-dispersive medium where the group velocity of light equals its phase velocity, the electromagnetic momentum of a single photon is $hf/(\sqrt{\varepsilon\mu}\,c)$; under these circumstances, the photon's mechanical momentum will be $[(\varepsilon+\mu-2)/(2\sqrt{\varepsilon\mu})]hf/c$.

**Example 2**. With reference to Fig. 2, consider a collimated, monochromatic beam of light propagating along the $z$-axis within a linear, isotropic, and homogeneous medium specified by its $(\varepsilon, \mu)$ parameters. The beam, which has a finite-diameter along the $x$–axis and an infinite diameter along $y$, is linearly polarized, having $E$-field amplitude $E_o\hat{\boldsymbol{x}}$ and $H$-field amplitude $H_o\hat{\boldsymbol{y}} = Z_o^{-1}\sqrt{\varepsilon/\mu}\,E_o\hat{\boldsymbol{y}}$ at the center. In addition, there exists a weak $E_z$ component of the field, which is an odd function of $x$ and goes to zero at the center. At the central $yz$-plane, the time-averaged rate of flow of $x$-momentum along the $x$-axis is given by Eq. (10a), as follows:

$$\langle T_{xx}\rangle = \langle -\tfrac{1}{2}\varepsilon_o^{-1}D_x^2 + \tfrac{1}{2}\mu_o H_y^2\rangle = \tfrac{1}{4}\varepsilon_o[(\varepsilon/\mu) - \varepsilon^2]E_o^2. \qquad (14)$$

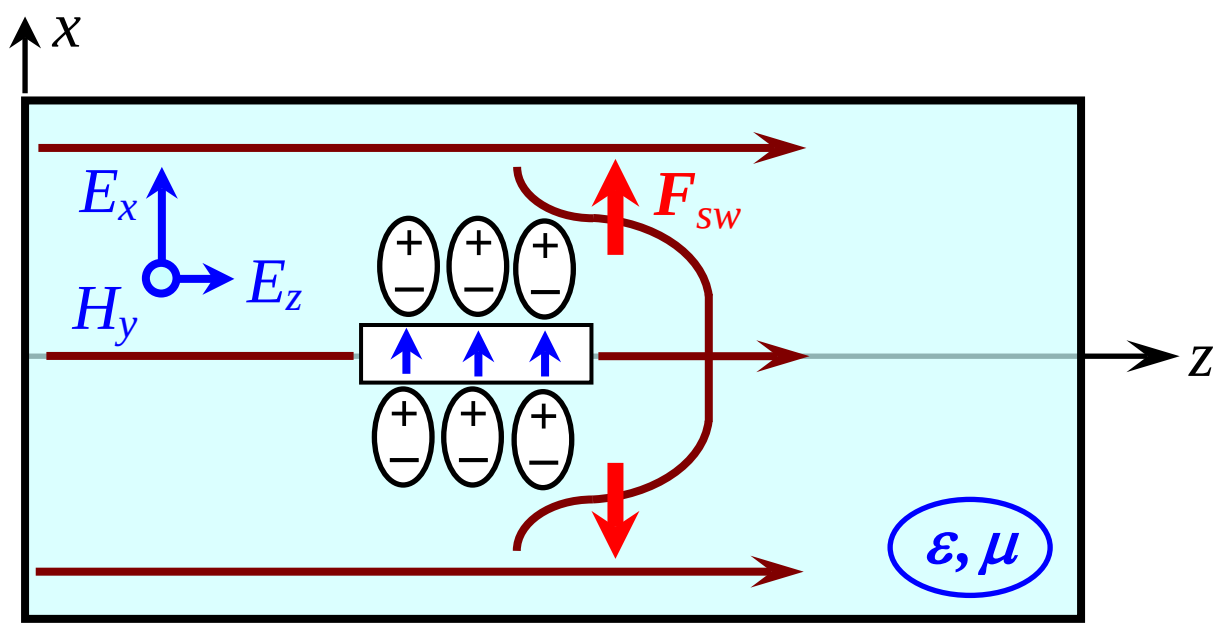

Fig. 2. A collimated beam having a finite diameter along the $x$-axis, propagates along $z$ in an isotropic, homogeneous medium specified by its $(\varepsilon, \mu)$ parameters. The beam has transverse magnetic (TM) or p-polarization, that is, its electromagnetic field components are $(E_x, E_z, H_y)$. A narrow gap opened in the central region of the beam reveals the existence of a force on the adjacent layers of dipoles. Continuity of $\boldsymbol{D}_\perp$ yields the $E$-field within the gap as $\varepsilon E_o\hat{\boldsymbol{x}}$. The $E$-field acting on the negative charges of the upper layer of the dipoles (as well as that acting on the positive charges of the lower dipoles) is $\tfrac{1}{2}(\varepsilon+1)E_o\hat{\boldsymbol{x}}$, whereas the field acting on the positive charges of the upper dipoles (or negative charges of the lower dipoles) is $E_o\hat{\boldsymbol{x}}$. These boundary dipole layers, therefore, experience an $E$-field gradient proportional to $\tfrac{1}{2}(\varepsilon-1)E_o\hat{\boldsymbol{x}}$. The net force of the $E$-field gradient exerted on the upper boundary layer is downward, while that on the lower boundary layers is upward. The two forces, being equal in magnitude, cancel each other out, but each force must be taken into account when considering the total force on the upper or lower halves of the medium. In addition to forces at the boundary layers, the sidewalls of the beam exert a force on the medium as well; the density of this force (per unit area of the sidewall) is denoted by $\boldsymbol{F}_{sw}$.

This momentum flow is entirely converted to a force on the electric dipoles located just above the $z$-axis and a second force, $\boldsymbol{F}_{sw}$, exerted on the medium by the upper sidewall of the beam. (The system being symmetric with respect to the $yz$-plane, identical forces, albeit in opposite directions, act on the lower half of the medium.) The force on the dipoles immediately above the $z$-axis is best understood if one introduces a gap in the middle of the beam as indicated in Fig. 2. The continuity of $\boldsymbol{D}_\perp$ at this interface reveals the $E$-field within the gap as being equal to $\varepsilon E_o\hat{\boldsymbol{x}}$. The average $E$-field at the interface is thus $\tfrac{1}{2}(\varepsilon+1)E_o$, and the field gradient sensed by the interfacial dipole layer is proportional to $\tfrac{1}{2}(\varepsilon-1)E_o$. The dipole density being $\boldsymbol{P} = \varepsilon_o(\varepsilon-1)E_o\hat{\boldsymbol{x}}$, we find a force density at the interface given by $\langle F_x\rangle = \tfrac{1}{4}\varepsilon_o(\varepsilon-1)^2E_o^2$. Adding this force density to $\langle T_{xx}\rangle$ of Eq. (14) yields $\langle F_x^{(sw)}\rangle = \tfrac{1}{4}\varepsilon_o[(\varepsilon/\mu) - 2\varepsilon + 1]E_o^2$, which is consistent with the sidewall force density of finite-diameter beams found in [3].

**Example 3**. Figure 3 shows a collimated, monochromatic beam of finite width propagating in a homogeneous medium specified by its $(\varepsilon, \mu)$ parameters. The propagation direction makes an angle $\theta$ with the $z$-axis in the $xz$-plane. The stress tensor of Eq. (10) gives the following time-averaged rate of flow of momentum (per unit area per unit time) across the $xy$-plane:

$$\langle T_{xz}\hat{\boldsymbol{x}} + T_{zz}\hat{\boldsymbol{z}}\rangle = -\langle E_x D_z\rangle\hat{\boldsymbol{x}} + ½\langle\varepsilon_o E_x^2 - \varepsilon_o^{-1} D_z^2 + \mu_o H_y^2\rangle\hat{\boldsymbol{z}}$$

$$= ½\varepsilon_o E_o^2\{\varepsilon\cos\theta\sin\theta\hat{\boldsymbol{x}} + ½[\cos^2\theta - \varepsilon^2\sin^2\theta + (\varepsilon/\mu)]\hat{\boldsymbol{z}}\}. \quad (15)$$

From Eq. (12), the rate of flow of momentum in a beam of cross-section $\cos\theta$ should be

$$\mathrm{d}\langle\boldsymbol{p}\rangle/\mathrm{d}t = ¼\,\varepsilon_o[1 + (\varepsilon/\mu)]E_o^2\cos\theta(\sin\theta\hat{\boldsymbol{x}} + \cos\theta\hat{\boldsymbol{z}}). \quad (16)$$

From Example 2, we find the imbalance of the sidewall force on the lower sidewall depicted in Fig. 3 to be

$$\langle\boldsymbol{F}_{sw}\rangle = ¼\,\varepsilon_o[(\varepsilon/\mu) - 2\varepsilon + 1]E_o^2\sin\theta(-\cos\theta\hat{\boldsymbol{x}} + \sin\theta\hat{\boldsymbol{z}}). \quad (17)$$

A narrow gap, opened parallel to the $x$-axis, reveals the force exerted on the boundary electric dipoles due to the $E_z$ discontinuity. The strength of this dipole layer is $\boldsymbol{P} = \varepsilon_o(\varepsilon - 1)E_o\sin\theta\hat{\boldsymbol{z}}$, and the effective $E$-field gradient acting on it is proportional to $½(\varepsilon - 1)E_o\sin\theta\hat{\boldsymbol{z}}$. The effective force on the dipole layer (per unit area) is thus given by

$$\langle\boldsymbol{F}\rangle = -¼\,\varepsilon_o(\varepsilon - 1)^2 E_o^2\sin^2\theta\hat{\boldsymbol{z}}. \quad (18)$$

The combined forces in Eqs. (16-18) are identical with the momentum flow rate of Eq. (15). The stress tensor of Eq. (10) is thus seen to yield the correct rate of momentum flow along the propagation direction when the relevant boundary forces are properly taken into account.

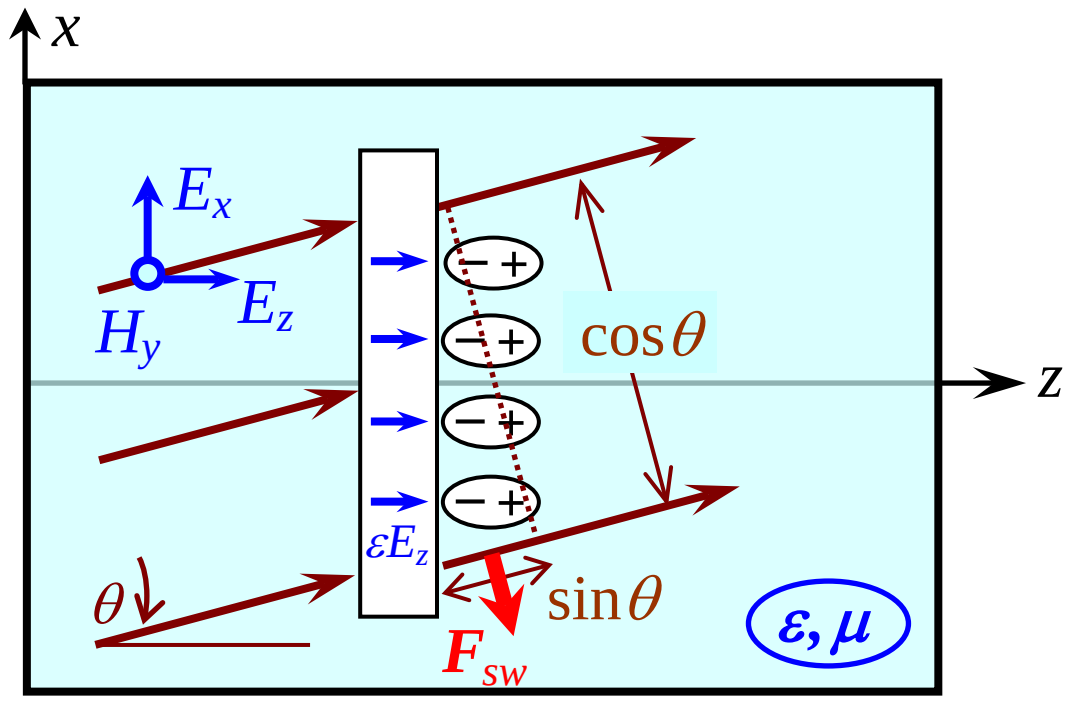

Fig. 3. A collimated beam of finite diameter in the $x$-direction (and infinite diameter along $y$) propagates in a medium specified by its $(\varepsilon, \mu)$ parameters. The propagation direction makes an angle $\theta$ with to the $z$-axis in the $xz$-plane. The beam's foot-print on the $x$-axis has unit length, making the beam width equal to $\cos\theta$, as shown. The beam is transverse magnetic (TM) or p-polarized, that is, its electromagnetic field components are $(E_x, E_z, H_y)$. A narrow gap, opened parallel to the $x$-axis, reveals the force exerted on the boundary layer electric dipoles due to $E_z$ discontinuity. The effective $E$-field gradient acting on the boundary dipole layer is proportional to $½(\varepsilon - 1)E_o\sin\theta\hat{\boldsymbol{z}}$. There is also an imbalance between the forces acting at the beam's upper and lower sidewalls, due to the extra length $\sin\theta$ of the lower wall. When the force on the boundary dipole layer as well as the imbalance of the sidewall forces are taken into account, the stress tensor component $T_{xz}\hat{\boldsymbol{x}} + T_{zz}\hat{\boldsymbol{z}}$ yields the rate of flow of momentum along the propagation direction.

## 3. Concluding remarks

We have derived the electromagnetic stress tensor of Eq. (10) for ponderable media in terms of the macroscopic fields $\boldsymbol{E}$, $\boldsymbol{D}$, $\boldsymbol{H}$, and $\boldsymbol{B}$. When the forces acting at the boundaries of a specified region of the medium are properly taken into account, we have shown that the stress tensor, in accordance with Eq. (11), yields the correct rate of flow of momentum, including both electromagnetic and mechanical momenta.

## Acknowledgements

This work is supported by the Air Force Office of Scientific Research under contract number FA 9550–04–1–0213. The author is grateful to Ewan M. Wright for helpful discussions.